\documentclass[a4paper,11pt]{article}
\usepackage{graphicx}
\newcommand{\del}{\partial}

\title{Perturbative analysis on
      infrared and ultraviolet aspects \\
       of noncommutative QED on $R^4$}
\author{M. Hayakawa
        \thanks{electric address: haya@post.kek.jp} \\
        {\it Theory Division, KEK,
             Tsukuba, Ibaraki 305-0801, Japan}   
       }
\date{\today}
\begin{document}
\maketitle
\begin{abstract}
 Here we discuss the ultraviolet and infrared
aspects of the noncommutative counterpart
of QED, which is called as noncommutative QED,
as well as some infrared dynamics of
noncommutative Yang-Mills (NCYM) theory.
 First we demonstrate
that the divergence in the theory
can be subtracted by the similar counterterms
as in ordinary theory at one loop level.
 Then the anomalous magnetic moment is calculated
to see the infrared aspect of the theory
which reflects the violation of Lorentz symmetry.
 The evaluation of the finite part
of the photon vacuum polarization
shows
that the logarithmically singular term
in the infrared limit
appears with the same weight as UV logarithmic divergence,
showing the correlation between the UV and infrared dynamics
in NCYM theory.
 NC-QED theory does not show such a property.
 We also consider the extension to chiral gauge theory
in the present context,
but the requirement of anomaly cancellation
allows only noncommutative QED.
\end{abstract}
%%%
%%%%%%%%%%%%%%%%%%%%%%%%%%%%%%%%%%%%%%%%%%%%%%%%%%%%%%%%%%
%%%%%%%%%%%%%%%%%%%%%%%%%%%%%%%%%%%%%%%%%%%%%%%%%%%%%%%%%%
%%%%%%%%%%%%%%%%%%%%%%%%%%%%%%%%%%%%%%%%%%%%%%%%%%%%%%%%%%
\section{Introduction}
\label{sec:intro}
%%%%%%%%%%%%%%%%%%%%%%%%%%%%%%%%%%%%%%%%%%%%%%%%%%%%%%%%%%
\quad
 The analysis of quantum mechanical features
of noncommutative field theory
is now being developed
from purely field theoretical point of view
[1--12].
%%\cite{Okawa}--\cite{Hayakawa}.
%%
%%\cite{Okawa,Filk,Martin_NCYM,J_NCYM,Krajewski,Bigatti,
%%Maldacena_NCYM,Alishahiha_NCYM,Ishibashi_NCYM,Seiberg,
%%Arefeva,Hayakawa}.
 The common character of
noncommutative field theories is its nonlocality.
 The product of the operators on the noncommutative geometry
is mapped into the stared (referred to hereafter as $*$-) product
on the deformed algebra of the usual functions.
 This gives the above momentum-dependent phase factor
for each interaction vertex which manifests nonlocality
of the theory.
 This phase factor selects
the diagrams with the ultraviolet (UV) divergence
since it can serve as an effective cutoff of high frequency modes
\cite{Filk,Bigatti,Ishibashi_NCYM,Seiberg,Hayakawa}.
 As a consequence
the UV-divergent diagrams
correspond to the 'planar' diagrams;
no phase suppression factors in the noncommutative theory side;
the 't Hooft diagrams \cite{tHooft}
which can be drawn on a plane
in the ordinary large N SU(N) gauge theory side
(a product of plaquette variables
 with no $Z_N$-phase factors
 in the twisted reduced model \cite{Okawa}).
 Thus noncommutative Yang-Mills (NCYM) theory
would be equivalent to the large N ordinary Yang-Mills system
at least in the high momentum region
as argued at the level of Feynman diagram
\cite{Bigatti,Ishibashi_NCYM}.
\\
\quad
 The paper intends to describe
the ultraviolet aspects of noncommutative QED
(NC-QED) in Sec. \ref{sec:NCQED}
which has not been reported in short article \cite{Hayakawa}
as well as the detail about the computation
of the infrared aspects common to NCYM theory
in Sec. \ref{sec:infrared} and Appendix.
 To make this paper self-contained
and explain the results more clearly,
the other parts of Ref. \cite{Hayakawa} should also be described
in detail.
 It is expected to
guide and help one to further study
the quantum mechanical aspect of noncommutative field theory.
\\
\quad
 Thus the paper is organized as follows:
Sec. \ref{sec:basic} is concerned with 
incorporation of the matter fields and showing
that the allowed choice is quite limited.
 In Sec. \ref{sec:NCQED} NC-QED theory
is quantized to investigate ultraviolet aspect
to demonstrate that
the one-loop divergence
can be subtracted by the usual set of local counterterms.
 Sec. \ref{sec:infrared} studies the infrared aspects
of the theory through
the anomalous magnetic dipole moment and
vacuum polarization of photon.
 The extension to the chiral gauge theory is also
examined in Sec. \ref{sec:chiral},
but it is found that there is {\it no} chiral gauge theory.
 Sec. \ref{sec:conc} is devoted to
the summary of the present paper.
%
%%%%%%%%%%%%%%%%%%%%%%%%%%%%%%%%%%%%%%%%%%%%%%%%%%%%%%%%%%
\section{Construction of Classical Action}
\label{sec:basic}
%%%%%%%%%%%%%%%%%%%%%%%%%%%%%%%%%%%%%%%%%%%%%%%%%%%%%%%%%%
\quad
 We begin with pure U(1) NCYM system,
which has the classical action in space-time dimension $d$
\begin{equation}
 S_{YM} = \int d^d x\,
          \left( -\frac{1}{4 g^2} \right)
           F_{\mu\nu} *F^{\mu\nu}
          \, .
 \label{eq:NYM_action}
\end{equation}
 Here the field strength $F_{\mu\nu}$ is
\begin{equation}
 F_{\mu\nu} = \del_\mu A_\nu - \del_\nu A_\mu
              -i [A_\mu,\, A_\nu]_{\rm M}\, ,
\end{equation}
where $[A,\, B]_{\rm M}$ denotes Moyal bracket:
\begin{equation}
 [A,\, B]_{\rm M} = A * B - B * A\, .
  \label{eq:Moyal}
\end{equation}
 The $*$-product
\begin{equation}
 A * B(x) \equiv
 \left.
  e^{\frac{1}{2i} C^{\mu\nu}
     \del^{(\xi)}_\mu \del^{(\eta)}_\nu}\,
  A(x+\xi)\, B(x+\eta)
 \right|_{\xi,\eta \rightarrow 0} \, ,
\end{equation}
is defined in terms of
the infinite towers of derivative operation
with use of an antisymmetric matrix $C^{\mu\nu}$
($C^{\mu\nu}$ has the dimension of area.)
which reflects noncommutativity of space-time
by modifying the algebra of functions.
 Even in U(1) case $A_\mu$ couples to itself
since the field strength $F_{\mu\nu}$
has the nonlinear term in $A_\mu$.
 The $*$-product obeys the associative law
which is also satisfied by the matrix algebra,
 Thus the algebraic manipulation in $*$-product
is the same as in the calculus of matrix.
 Then, assuming that the fields decrease so promptly at infinity
that the space-time integral of a Moyal bracket
(which corresponds to the trace of the commutator
in the ``matrix language'') vanishes,
it is easily shown that
the action (\ref{eq:NYM_action}) is invariant under
the gauge transformation
\begin{equation}
 A_\mu(x) \rightarrow A^\prime(x)
  = U(x) *A_\mu(x) *U^{-1}(x) + i U(x) *\del_\mu U^{-1}(x)\, ,
  \label{eq:gauge_A}
\end{equation}
where $U(x) = (e^{i\theta(x)})_*$ is defined
by an infinite series of multiple $*$-product of
scalar function $\theta(x)$.
 The the similarly defined $U^{-1}(x) = (e^{-i\theta(x)})_*$
corresponds to the inverse of $U(x)$.
%%%%%
\\
\quad
 The coupling of ``electron'' to gauge field
in U(1) NCYM theory
receives a severe restriction.
 Here we mean by ''electron'',
the fields which couple to the gauge fields
in the similar manner as in ordinary QED,
in which the covariant derivative for the matter field
$\psi$ with charge $Q_\psi$ is given by
\begin{equation}
 D_\mu \psi = \del_\mu \psi - iQ_\psi A_\mu \psi\, .
 \label{eq:o_cov}
\end{equation}
 Conventional assignment of charge $Q_\psi$
in the electroweak theory is like
$Q_e=-1$ for electron, $Q_u=\frac{2}{3}$ for up-type quark,
etc.
 As is well-known there is no reason for $Q_\psi$
to be quantized in the electroweak theory.
 Extension to noncommutative case
shows that nonzero U(1) charge is not only quantized,
but also limited to be $+1$ or $-1$
even on Minkowski space-time.
 This contrasts with the mechanism of charge quantization
on compact manifold from the single-valuedness
of the wave function.
 We show those facets below.
\\
\quad
 First a simple algebraic manipulation shows that
a combination
\begin{equation}
 D_\mu \psi = \del_\mu \psi - i A_\mu *\psi \, ,
\end{equation}
which resembles (\ref{eq:o_cov}),
behaves covariantly
\begin{equation}
 D_\mu \psi(x) \rightarrow
  D^\prime_\mu \psi^\prime(x) = U(x) *D_\mu \psi(x)\, ,
\end{equation}
under transformation (\ref{eq:gauge_A}) for $A_\mu(x)$
and
\begin{equation}
 \psi(x) \rightarrow
  \psi^\prime(x) = U(x) *\psi(x)\, ,
\end{equation}
for $\psi(x)$.
 Likewise if $\hat{\psi}(x)$ transforms as
\begin{equation}
 \hat{\psi}(x) \rightarrow
  \hat{\psi}^\prime(x) = \hat{\psi}(x) *U^{-1}(x)\, ,
 \label{eq:gauge_h_psi}
\end{equation}
then the quantity
\begin{equation}
 D_\mu \hat{\psi} = \del_\mu \hat{\psi}
                     + i \hat{\psi} *A_\mu\, ,
\end{equation}
behaves in the same manner as (\ref{eq:gauge_h_psi}).
 Commutative limit $C^{\mu\nu} \rightarrow 0$
indicates that $\hat{\psi}$ corresponds
to a field in ordinary U(1) gauge theory
with the same magnitude but opposite sign
compared to the one associated with $\psi$.
 Thus we say that $\hat{\psi}$ has charge $-1$
while $\psi$ carries $+1$.
\\
\quad
 Therefore, for instance, the action
%%%
\begin{equation}
 S_{\rm matter} = \int d^d x
  \left(
   \bar{\psi} *\gamma^\mu iD_\mu \psi - m \bar{\psi} *\psi
  \right)
  \, ,
   \label{eq:matter_action}
\end{equation}
has local U(1) invariance
since $\bar{\psi}$ transforms as $\hat{\psi}$.
 However a simple exercise shows that the simple extension
\begin{equation}
 D_\mu \psi^{(n)} = \del_\mu \psi^{(n)}
                    - i n A_\mu *\psi^{(n)}\, ,
\end{equation}
with $\psi^{(n)} \rightarrow
      \psi^{(n)\,\prime} = U^n \psi^{(n)}$ 
for the field $\psi$ with integral multiple $n$ of unit charge
fails to transform covariantly.
\\
\quad
 Derivation of the above facts and the various formula
is similar to that in ordinary non-abelian
gauge theory due to the simple fact that
$*$-product satisfies the associative law.
 The field with charge $+1$($-1$) in noncommutative case
would correspond to (anti-)fundamental representation
in ordinary nonabelian gauge theory.
 It is also reminiscent of such features that
noncommutative gauge theory carries
the internal degrees of freedom by imbedding them
into the space-time geometry itself.
 This is the reverse process of the reduction
of the space-time degrees of freedom
into the internal ones in the large N gauge theory
\cite{Kawai_RM,Okawa}.
 When we pursue this correspondence further,
we are inclined to guess that
the higher-rank representation of SU(N) gauge theory
may convert into some matter fields
in noncommutative gauge theory.
 It would be the counterpart of the fields
with an integral multiple of unit charge
from the view point of
noncommutative generalization of U(1) gauge theory.
 Actually the adjoint representation
corresponds to a field $\chi(x)$ with zero charge in total
but transforming in the by-product form
\begin{equation}
 \chi(x) \rightarrow \chi^\prime(x) =
                      U(x) *\chi(x) *U^{-1}(x)\, .
\end{equation}
 Its covariant derivative is given by Moyal bracket
(\ref{eq:Moyal}).
 What is the counterpart,
for instance, of the second-rank antisymmetric representation
of SU(N) gauge theory ?
 We could not succeed to find those other counterparts.
 Although a consequence depends on groping the possibility,
we see that
the fields of nonzero charge with different absolute magnitude
cannot coexist in a system.
\\
\quad
 Note that the above argument
also persists for the static electron.
 Thus the the vacuum expectation value of Wilson loop
along the simple rectangular loop
is associated with the ground energy acting between
the static sources with charges $\pm 1$
similarly as in ordinary Yang-Mills theory.
%
%%%%%%%%%%%%%%%%%%%%%%%%%%%%%%%%%%%%%%%%%%%%%%%%%%%%%%%%%%%%%%%%
\section{Perturbation Theory of Noncommutative QED}
\label{sec:NCQED}
\begin{figure}[t]
\begin{center}
 \includegraphics[scale=0.5]{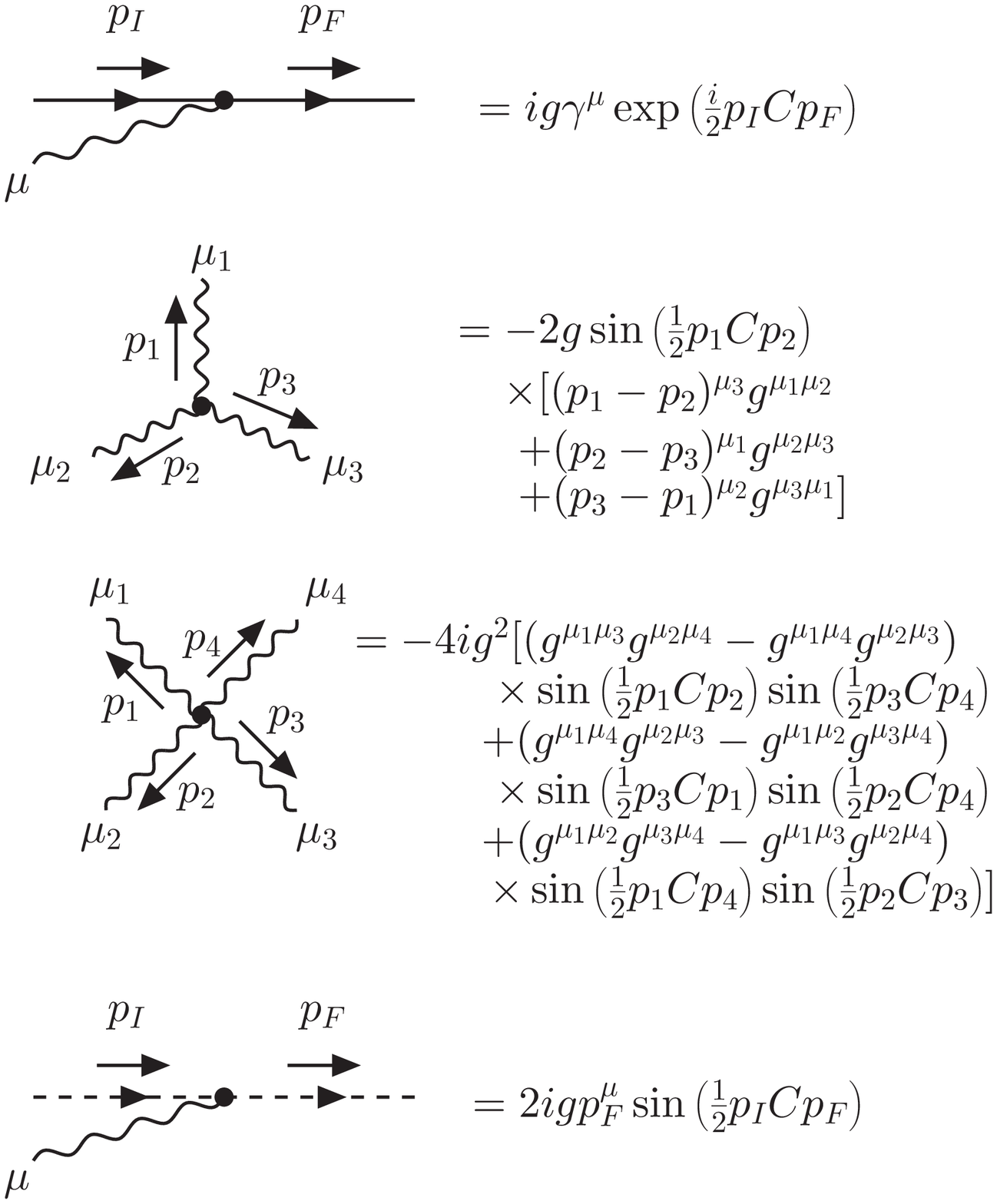}
 \caption{Feynman rule in noncommutative QED}
 \label{fig:Feynman_rule}
\end{center}
\end{figure}

%%%%%%%%%%%%%%%%%%%%%%%%%%%%%%%%%%%%%%%%%%%%%%%%%%%%%%%%%%%%%%%%
%
 To perform perturbation theory for NC-QED,
we rescale the gauge field appearing in the previous section
as $A_\mu \rightarrow g A_\mu$.
 To perform gauge fixing
to obtain the nonsingular free propagator
for the gauge fields,
BRST quantization similar to Ref. \cite{Martin_NCYM}
is adequate.
 Incorporating the ghost fields $c$, $\bar{c}$
and the auxiliary fields $B$,
BRST transformation
\begin{eqnarray}
 &&
 \hat{\delta}_{\bf B} A_\mu = D_\mu c
  = \del_\mu c - i g [A_\mu,\, c]_{\rm M}\, ,
  \nonumber \\
 &&
 \hat{\delta}_{\bf B} c= i\,g\,c*c\, ,\quad
 \hat{\delta}_{\bf B} \bar{c} = i B\, ,\quad
 \hat{\delta}_{\bf B} B = 0\,
  \nonumber \\
 &&
 \hat{\delta}_{\bf B} \psi = i c *\psi\, ,\quad
 \hat{\delta}_{\bf B} \bar{\psi} = i \bar{\psi} *c\,
\end{eqnarray}
is nilpotent.
 Under this transformation
the quantities in (\ref{eq:NYM_action})
and (\ref{eq:matter_action}) are BRST-closed respectively.
 The Faddeev-Popov and gauge fixing term
is introduced as BRST-exact form
\begin{equation}
 S_{\rm GF} = \int d^d x
              (-i \hat{\delta}_{\bf B})
              \frac{1}{2}
              \left(
               \bar{c} *\left(
                         \frac{\alpha}{2} B + \del_\mu A^\mu
                        \right)
               +
               \left(
                \frac{\alpha}{2} B + \del_\mu A^\mu
               \right) *\bar{c}
              \right)\, .
\end{equation}
 After elimination of $B$ through its equation of motion
the above quantity reduces to
\begin{equation}
 S_{\rm GF} =
  \int d^d x
  \left(
   -\frac{1}{2\alpha} \del_\mu A^\mu *\del_\nu A^\nu
   +
   \frac{1}{2}
   \left(
    i\bar{c} *\del^\mu D_\mu c -
    i\del^\mu D_\mu c * \bar{c}
   \right)
  \right)\, .
   \label{eq:GF_term}
\end{equation}
\\
\quad
 We consider NC-QED on Minkowski space-time
to estimate the on-shell electron coupling to
the photon.
 To simplify the discussion $*$-product works effectively
on one plane in the space in the canonical basis of $C^{\mu\nu}$.
 This is, we consider the case when
$C^{01} = 0$ but nonzero $C^{23}$.
 Then there are two kinds of on shell photon;
one with momentum along $(2,3)$-plane;
one with that moving orthogonal to $(2,3)$-plane.
 Thus we can discuss the effect of explicit Lorentz
invariance introduced by $C^{23}$.
\\
\quad
 As in Ref. \cite{Martin_NCYM}
quantization of noncommutative theory is realized
as a perturbative expansion by giving Feynman rules through
\begin{equation}
 Z[J] = \int D\Phi\,
         e^{iS[\Phi] + i\int d^d x J(x) *\Phi(x)}\, ,
\end{equation}
where the action $S$ is the sum of (\ref{eq:NYM_action}),
(\ref{eq:matter_action})
\begin{equation}
 S_{\rm NC-QED}
  = \int d^d x\,
    \left(
     -\frac{1}{4 g^2} F_{\mu\nu} *F^{\mu\nu}
     +
     \bar{\psi} *\gamma^\mu iD_\mu \psi - m \bar{\psi} *\psi
   \right)
  \, .
 \label{eq:NC-QED}
\end{equation}
together with (\ref{eq:GF_term}).
 As a result
the propagator is the same as in its commutative counterpart,
while each vertex accompanies with a phase factor
shown in Fig. \ref{fig:Feynman_rule}
depending on the momenta outgoing from the vertex.
 It manifests the nonlocal nature of theory.
%%%%
%-------------------------------------------------------------
%-------------------------------------------------------------
%%%%
%
\\
\quad
 In order to pursue ultraviolet (UV) divergent structure,
we begin with discussing a few typical vertex functions
(contribution to Green functions from
one particle irreducible Feynman diagrams
with all the external legs amputed)
in the succeeding subsections.
 This would be the prompt approach to learn
about the nature of our theory.
 The aim of our analysis below
is to demonstrate that
at one-loop level the present theory can be
made finite by redefining
the fields and parameters consistently
with the symmetry of the classical action.
  The analysis also shows that
UV divergence can appear only
in a restricted set of diagrams, ``planar diagram''
which will be defined at the end of Sec. \ref{sec:e_g_coupling}.
%
%%%%%%%%%%%%%%%%%%%%%%%%%%%%%%%%%%%%%%%%%%%%%%%%%%%%%%%%%%%%%%%%
\subsection{Two point functions}
\label{sec:two_point}
%%%%%%%%%%%%%%%%%%%%%%%%%%%%%%%%%%%%%%%%%%%%%%%%%%%%%%%%%%%%%%%%
%
%%%%
%-------------------------------------------------------------
\begin{figure}[t]
\begin{center}
 \includegraphics[scale=0.5]{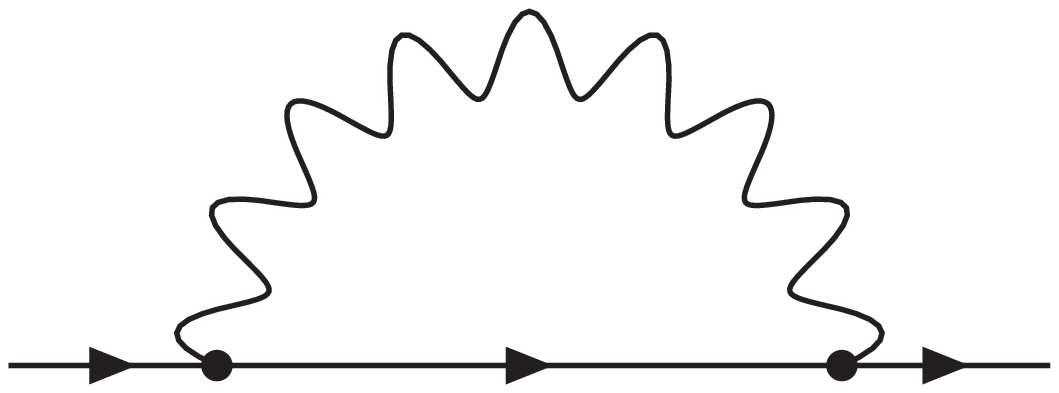}
 \caption{Correction to electron self-energy}
 \label{fig:electron}
\end{center}
\end{figure}
%-------------------------------------------------------------
%%%%
\quad
 The electron self-energy receives one-loop correction
through only one diagram shown in Fig. \ref{fig:electron}
as in ordinary QED.
 In this diagram
the phase factors accompanied with the two vertices
cancel with each other.
 Thus the contribution is nothing but the one
found in ordinary QED.
 Thus not only UV divergence can be subtracted by
the usual rescaling of wave function and mass
of electron, but also the remained finite part does not change
as long as the same renormalization scheme
($\overline{{\rm MS}}$-scheme here) is applied.
 The wave function renormalization factor $Z_\psi$
for the electron is in particular found as
%%%%
\begin{equation}
 Z_\psi = 1 - \frac{g^2}{16\pi^2} \frac{1}{\varepsilon^\prime}
  \, ,
  \label{eq:Z_psi}
\end{equation}
%%%%
where
  $1/\varepsilon^\prime$
  = $1/\varepsilon$ + $\gamma_E - {\rm ln}(4\pi)$
for the space-time dimension $d= 4 - 2\varepsilon$.
 Here and hereafter Feynman gauge ($\alpha=1$) is taken
to simplify the expression.
\\
\quad
%%%%
%-------------------------------------------------------------
\begin{figure}[t]
\begin{center}
 \includegraphics[scale=0.5]{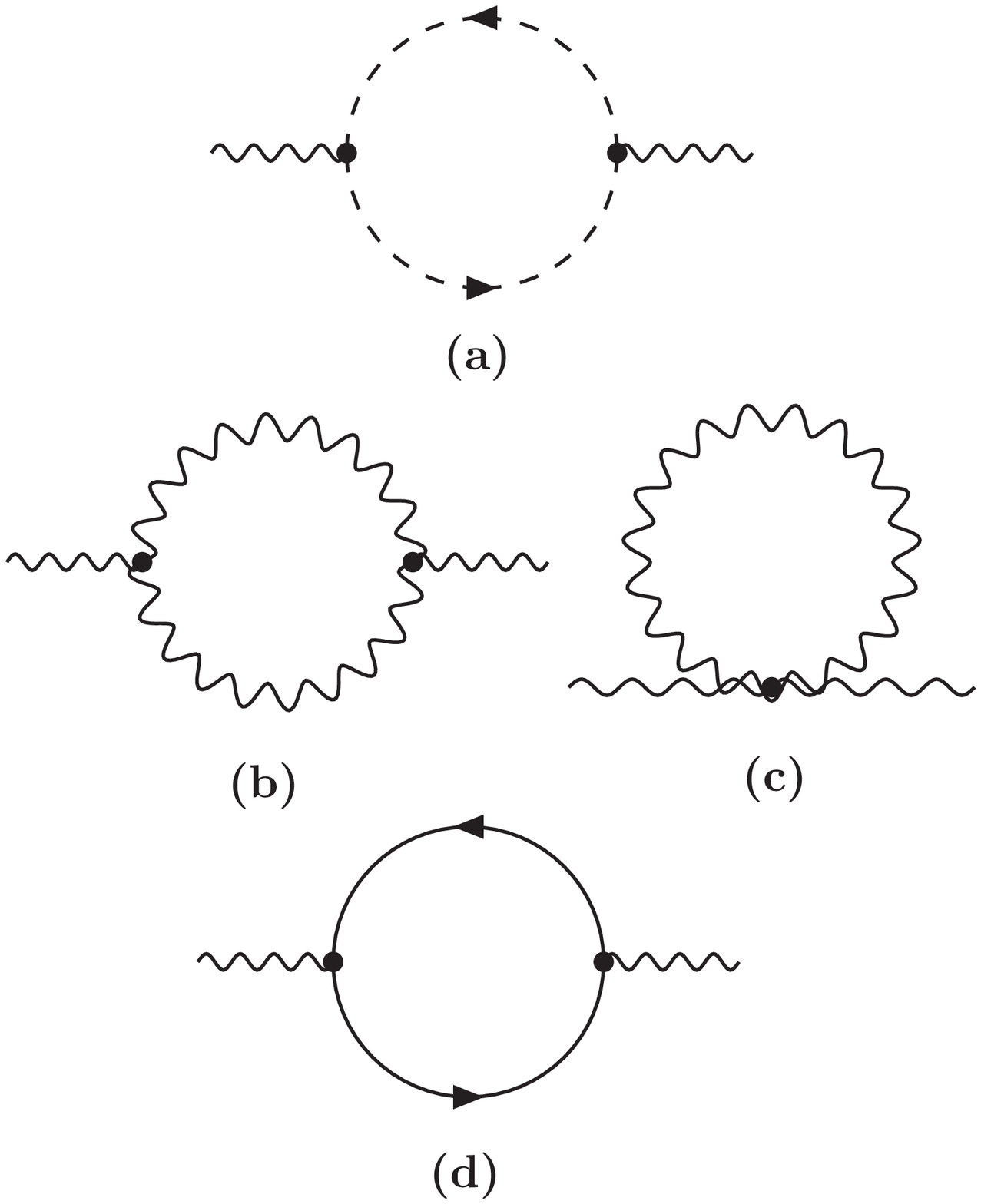}
 \caption{Correction to vacuum polarization of photon}
 \label{fig:vac_photon}
\end{center}
\end{figure}
%-------------------------------------------------------------
%%%%
 The photon self-energy diagram receives
an electron loop contribution (Fig. \ref{fig:vac_photon}(d))
as well as the FP ghost (Fig. \ref{fig:vac_photon}(a))
and gauge boson (Fig. \ref{fig:vac_photon}(b), (c)) loops
which already appear in NCYM theory.
 The evaluation of the diagram
is illustrated
for the ghost loop diagram in Fig. \ref{fig:vac_photon}(a)
in Appendix.
 Here at first we take up the simpler diagram,
the gauge boson diagram
in Fig. \ref{fig:vac_photon}(c)
\begin{equation}
 i\Pi^{\mu\nu}_{(c)}(q) =
 2(d-1) i g^2 g^{\mu\nu}
 \int_0^\infty id\alpha \frac{1}{[4\pi\alpha i]^{d/2}}
 \left(
  1 - \exp\left[ -i \frac{1}{\alpha}
                    \frac{\tilde{q}^2}{4}
          \right]
 \right)\, .
  \label{eq:VP_4_tent}
\end{equation}
 In dimensional regularization
the part which does not depend on any phase factor
vanishes (no divergence).
 On the other hand,
the portion of (\ref{eq:VP_4_tent})
arising from the contribution dependent on a loop momentum
(nonplanar contribution)
evades the integral to diverge
due to an effective damping factor,
$\exp\left( -i \frac{1}{\alpha}
               \frac{\tilde{q}^2}{4} \right) $.
 Thus it is finite
\begin{equation}
 i\Pi^{\mu\nu}_{(c)}(q) = 
  i \frac{g^2}{16\pi^2} g^{\mu\nu} \frac{-24}{-\tilde{q}^2}\, ,
   \label{eq:VP_4}
\end{equation}
which shows the $1/\tilde{q}^2$ singularity for
$\tilde{q}^2 \rightarrow 0$.
 It should be remembered that
such a portion of (\ref{eq:VP_4_tent})
that leads to (\ref{eq:VP_4}) became finite
due to finiteness of $\tilde{q}^2$.
 But it would lead ultraviolet singularity
if it were evaluated by setting $\tilde{q}^2$ equal to zero
in eq. (\ref{eq:VP_4_tent}).
 The above $1/\tilde{q}^2$-singularity
corresponds to quadratic UV divergence.
 However we know that quadratic UV divergence does
not appear in the gauge theory.
 As will be shown in next section,
(\ref{eq:VP_4_tent}) is cancelled by
the singular terms involved
in the contributions from Fig. \ref{fig:vac_photon}(a) and (b).
 The full expression for these latter contributions is found as
%%%
\begin{eqnarray}
 &&
 i\Pi^{\mu\nu}_{\rm (a) + (b)}(q) =
 ig^2 \int_0^\infty id\alpha_+ \int_0^\infty id\alpha_-
  \frac{1}{(4\pi\beta i)^{d/2}}
  \exp\left[
        -i \frac{\alpha_+ \alpha_-}{\beta} (-q^2)
      \right]
   \nonumber \\
 && \qquad \quad \times
 \left[
   \left(
     1 - \exp\left[ -i \frac{1}{\beta} \frac{\tilde{q}^2}{4} \right]
   \right)
   \times
   \left\{
     g^{\mu\nu} \left(
                 (3d-4) i\frac{1}{\beta} +
                 \left(
                   5-2\frac{\alpha_+ \alpha_-}{\beta^2}
                 \right) q^2
                \right)
   \right.
 \right.
  \nonumber \\
 && \qquad \qquad \qquad \qquad \qquad \qquad \qquad \quad
 + \left.
    q^\mu q^\nu
     \left(
       (d-6) - 4(d-2) \frac{\alpha_+\alpha_-}{\beta^2}
     \right)
   \right\}
   \nonumber \\
 && \qquad \qquad \qquad
\left.
 + \exp\left[ -i \frac{1}{\beta} \frac{\tilde{q}^2}{4} \right]
   \times \frac{1}{\beta^2} \times
   \left\{
    - \frac{1}{2} g^{\mu\nu} \tilde{q}^2
    + (2-d) \tilde{q}^\mu \tilde{q}^\nu
   \right\}
 \right]\, .
  \label{eq:VP_g_33}
\end{eqnarray}
%%%
 The ultraviolet singularity in it is removed
by the wave function renormalization for photon
\cite{Martin_NCYM}
%%%
\begin{equation}
 \left. Z_A \right|_{{\rm NCYM}} =
  1 + \frac{g^2}{16\pi^2} \frac{10}{3}
      \frac{1}{\varepsilon^\prime}\, .
   \label{eq:Z_A_NCYM}
\end{equation}
%%%
 For an electron loop contribution
in Fig. \ref{fig:vac_photon}(d)
the phases cancel between the two vertices
as in the correction to electron self-energy.
 There is no change from ordinary QED
for this additional contribution.
 Combined with (\ref{eq:Z_A_NCYM})
the singularity in this contribution
gives for the wave function renormalization factor $Z_A$
in total
\begin{equation}
 Z_A = 1 +
       \frac{g^2}{16\pi^2}
        \left(
         \frac{10}{3} - \frac{4}{3} N_F
        \right) \frac{1}{\varepsilon^\prime}\, ,
  \label{eq:Z_A}
\end{equation}
where $N_F$ denotes the number of independent fields
with charge $\pm1$.
%%%%%%%%%%%%%%%%%%%%
\\
\quad
 The conclusion of this subsection is that
all the ultraviolet divergences
can be subtracted away by the same local counterterms
as in ordinary QED.
%%
%%%%%%%%%%%%%%%%%%%%%%%%%%%%%%%%%%%%%%%%%%%%%%%%%%%%%%%%%%%%%%%%
\subsection{Electron coupling to photon}
\label{sec:e_g_coupling}
%%%%%%%%%%%%%%%%%%%%%%%%%%%%%%%%%%%%%%%%%%%%%%%%%%%%%%%%%%%%%%%%
%%
%-------------------------------------------------------------
\begin{figure}[t]
\begin{center}
 \includegraphics[scale=0.5]{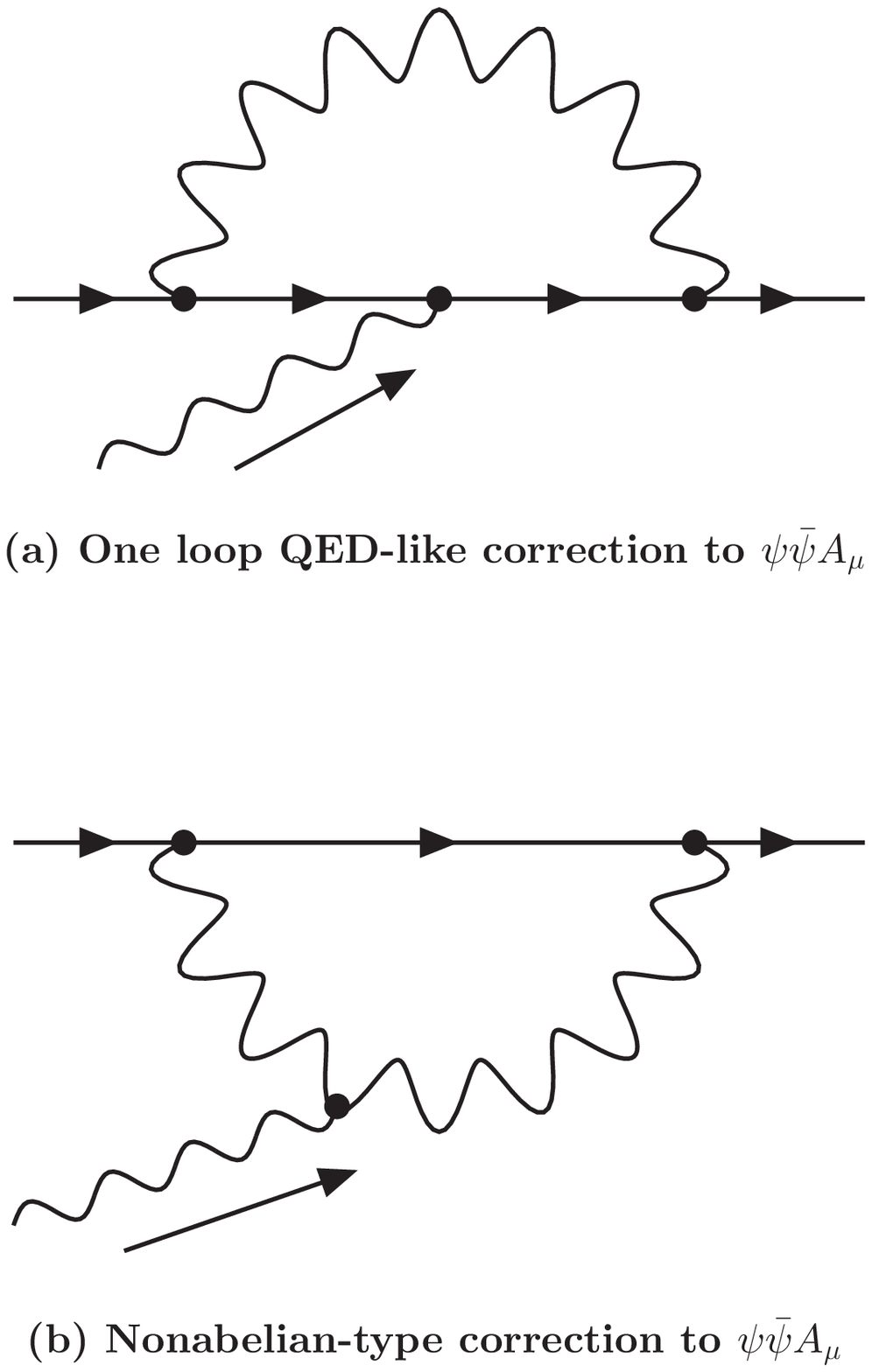}
 \caption{Correction to $\psi\bar{\psi}A_\mu$ vertex}
 \label{fig:e_g}
\end{center}
\end{figure}
%-------------------------------------------------------------
%%
\quad
 On account of finding the correction to
magnetic dipole moment in the next section
we describe the detail of one loop correction to
the interaction of the electron to photon mainly.
 There are two topologically distinct diagrams
which contribute to this interaction.
 The first one shown in  Fig. \ref{fig:e_g}(a)
is the analogue of QED
while the other occurs through a nonabelian vertex
involving three gauge bosons
as shown in Fig. \ref{fig:e_g}(b).
 The QED-like diagram in Fig. \ref{fig:e_g}(a)
is finite while the nonabelian-type diagram
in Fig. \ref{fig:e_g}(b) leads UV-divergence.
 The local counterterm with the coefficient
\begin{equation}
 Z_{\bar{\psi}\psi A} =
  1 - \frac{g^2}{16\pi^2} \frac{1}{\varepsilon^\prime}
       \times 3\, ,
  \label{eq:Z_e_g}
\end{equation}
is sufficient to remove this divergence.
%%%%%%
\par
 Here we would like to check
whether we identify UV divergence properly,
and quantize a system maintaining local gauge invariance.
 The nontrivial point to be examined here
is the universality of gauge coupling constant.
 This implies the relations among the rescale factors
$Z_\phi$ for each field $\phi$,
$Z_{\phi_1 \phi_2 \phi_3}$ for each coupling
to which $\phi_1$, $\phi_2$ and $\phi_3$ participate
\begin{equation}
 \frac{Z_{\bar{\psi} \psi A}}{Z_\psi}
 =
 \frac{Z_{AAA}}{Z_A}
 =
 \sqrt{\frac{Z_{AAAA}}{Z_A}}
 =
 \frac{Z_{\bar{c}cA}}{Z_c}\, .
  \label{eq:univ}
\end{equation}
 Note that
 $Z_{\bar{c}cA}$ and ${Z_c}$
do not alter even after incorporation of electrons
at one-loop level.
 Thus the final combination found in eq. (\ref{eq:univ})
is completely the same as in NCYM case
\begin{equation}
 \frac{Z_{\bar{c}cA}}{Z_c} =
  1 - \frac{g^2}{16\pi^2} \frac{1}{\varepsilon^\prime} \times 2
  \, .
   \label{eq:ccAc}
\end{equation}
 The relation (\ref{eq:univ}) and this fact
insist that the effects of incorporation of electron
must cancel between the denominator
and the numerator in each of the other three combinations
in (\ref{eq:univ}).
 Before observing this fact we recall the situation
in ordinary QED.
 In ordinary QED the relation (\ref{eq:univ}) reduces to
\begin{equation}
 \left. \frac{Z_{\bar{\psi}\psi A}}{Z_\psi} \right|_{\rm QED}
  = 1\, .
\end{equation}
 That is, $\left. Z_\psi \right|_{\rm QED}$
is required to balance completely with
$\left. Z_{\bar{\psi}\psi A} \right|_{\rm QED}$.
 $\left. Z_{\bar{\psi}\psi A} \right|_{\rm QED}$ is derived
from UV-divergence of the similar diagram
in Fig. \ref{fig:e_g}(a).
 Returning to NC-QED,
the argument in Sec. \ref{sec:two_point}
showed that $Z_\psi$ is
the same (\ref{eq:Z_psi}) found in ordinary QED,
$Z_\psi = \left. Z_\psi \right|_{\rm QED}$.
 However in NC-QED the QED-like diagram
in Fig. \ref{fig:e_g}(a) does not diverge
contrary to the case in QED.
(Note that this feature persists
for any choice of gauge fixing parameter.)
 Those facts raise the question
whether the relation (\ref{eq:univ}) is actually maintained or not.
 Somewhat curiously
$Z_{\bar{\psi}\psi A}$ in eq. (\ref{eq:Z_e_g})
obtained from  Fig. \ref{fig:e_g}(b) and $Z_\psi$
in (\ref{eq:Z_psi}) establishes eq. (\ref{eq:univ})
with (\ref{eq:ccAc}).
 The direct computation also shows
that all the divergences in $AAA$- and $AAAA$-vertices
can be subtracted by the usual countertems
and the associated $Z$ factors satisfy the relations
required in eq. (\ref{eq:univ}).
\\
\quad
 Also in the present theory,
the $\beta$ function signifies
whether the theory is well defined at an UV fixed point
or not.
 It also helps one to improve the accuracy of the approximation
by resuming the large logarithmic corrections.
 However we must always keep in mind that
the typical behavior at some fixed energy scale
cannot be drawn from the usual intuition based on
the block-spin renormalization group method
applicable to local field theory.
\\
\quad
 The beta function is the simplest to find
by observing the gauge interaction of Faddeev-Popov ghosts.
 There only $Z_A$ derived from the singularity of
the photon vacuum polarization receives the electron effect
at one-loop level.
 Eqs. (\ref{eq:ccAc}) and (\ref{eq:Z_A}) lead
\begin{equation}
 \beta(g) = \frac{1}{g} Q \frac{dg}{dQ}
  = -\left(
       \frac{22}{3} - \frac{4}{3} N_F
     \right)\, \frac{g^2}{16\pi^2}\, .
   \label{eq:beta}
\end{equation}
 A contribution $\frac{22}{3}$ is due to the structure similar
to nonabelian dynamics
which is pure SU(2) Yang-Mills theory
\cite{Martin_NCYM,J_NCYM}.
 However the matter contribution is that found in QED theory
with unit charge, {\it not} that of the quarks belonging to
the fundamental representation of SU(2) gauge theory
(where $\frac{2}{3}$ instead of $\frac{4}{3}$ per flavor
in (\ref{eq:beta})).
\\
\quad
 The above analysis indicates
the following general features about UV divergence.
 First we write a diagram according to Feynman rule
in Fig. \ref{fig:Feynman_rule}.
 Then the associated contributions
are divided into the pieces each of which
has a definite momentum-dependent phase factor.
 If a given piece admits a loop momentum
which does not appear in any phase factor,
it can lead divergence
(It may be a subdivergence which should be subtracted
by the counterterms determined at the previous order
of perturbation.)
 We define such a diagram as ``planar'' diagram.
 Power counting argument allows us to choose
the superficially divergent pieces among those planar diagrams.
 Since the same mechanism above works in the more complex
diagrams appearing
in the vertex functions with three or more external legs,
UV divergence can only appear
in the vertex functions corresponding to
that in the ordinary QCD (three gauge boson vertices, etc.).
 Determination of the above fact
due to 't Hooft-type representation may be developed
along the same line as done for NCYM theory in Ref.
\cite{Filk,Ishibashi_NCYM,Seiberg}.
% With inclusion of electron, there arises
%another kind of diagram,
%disk diagrams 
%which can be drawn on a plane but either
%lead divergence (e.g., Fig. \ref{fig:vac_photon}(d))
%or does not lead divergence (e.g., Fig. \ref{fig:e_g}(a))
%\footnote{
% This is suggested by S. Iso.
%}.
%%%
%%%%%%%%%%%%%%%%%%%%%%%%%%%%%%%%%%%%%%%%%%%%%%%%%%%%%%%%%%%%%%%%%
\section{Some Infrared Aspects}
\label{sec:infrared}
%%%%%%%%%%%%%%%%%%%%%%%%%%%%%%%%%%%%%%%%%%%%%%%%%%%%%%%%%%%%%%%%%
%%%
\quad
 We are more interested in the infrared aspect
under the situation that
UV behavior is shown to be greatly modified
due to nonabelian nature of noncommutative field theory.
 The small number of flavors insures
that the theory is asymptotically free.
 This behavior suggests the strong coupling low energy dynamics
of the theory.
 But here we assume that the coupling is kept small
and the theory admits the perturbative analysis.
 The perturbative infrared aspect is examined
by calculating the leading correction to magnetic dipole moment
($g-2$) in Sec. \ref{sec:MDM}
as well as
the vacuum polarization of the photon in Sec. \ref{sec:vac},
and ask what modification occurs in the extension to NC-QED.
%%
%%%%%%%%%%%%%%%%%%%%%%%%%%%%%%%%%%%%%%%%%%%%%%%%%%%%%%%%%%%%%%%%%
\subsection{Anomalous magnetic dipole moment}
\label{sec:MDM}
%%%%%%%%%%%%%%%%%%%%%%%%%%%%%%%%%%%%%%%%%%%%%%%%%%%%%%%%%%%%%%%%%
\quad
 The extraction of dipole coupling
from Figs. \ref{fig:e_g}(a) and \ref{fig:e_g}(b) yields
\begin{eqnarray}
 &&
 i g^3
 \left[
  e^{\frac{i}{2} p_I \cdot C \cdot p_F}\,H(1,p,q)
 \right.
  \nonumber \\
 && \quad \quad
  +
 \left.
  e^{\frac{i}{2} p_I \cdot C \cdot p_F}\,H(0,p,q)
  -
  e^{-\frac{i}{2}p_I \cdot C \cdot p_F}\,H(1,p,q)
 \right]\, mi\sigma^{\mu\nu} q_\nu \, ,
 \label{eq:MDM_1}
\end{eqnarray}
where $q$ is the incoming photon momentum,
and $p$ is connected to the incoming electron momentum
$p_I$ and the outgoing electron momentum $p_F$ through
\begin{equation}
 p_I = p - \frac{q}{2}, \quad p_F = p + \frac{q}{2}\, .
\end{equation}
 The matrix $\sigma^{\mu\nu}$ is here
$\sigma^{\mu\nu}
 = \frac{i}{2} \left[ \gamma^\mu, \gamma^\nu \right]$.
 The contribution in the first line in eq. (\ref{eq:MDM_1})
is from Fig. \ref{fig:e_g}(a) and
one in the second line from Fig. \ref{fig:e_g}(b).
 The function $H(\eta,p,q)$ appearing in (\ref{eq:MDM_1}) is
\begin{eqnarray}
 H(\eta,p,q) &=& \int_0^{\infty} id\alpha_0
  \int_0^{\infty} id\alpha_+ \int_0^{\infty} id\alpha_+
  \frac{1}{[4\pi\beta i]^2}
  \nonumber \\
  && \quad \quad
  \times 2
  \left(
   \frac{\alpha_+ + \alpha_-}{\beta}
   -
   \left(
    \frac{\alpha_+ + \alpha_-}{\beta}
   \right)^2
  \right)
   \nonumber \\
 && \quad \quad
   \times
   \exp \left[
         -i\frac{1}{\beta}
           \left\{
            (\alpha_+ + \alpha_-)^2 m^2
             + \alpha_+ \alpha_- (-q^2)
           \right.
        \right.
         \nonumber \\
 && \qquad \qquad \qquad \qquad
       \left.
           \left.
             - \eta (\alpha_+ + \alpha_-) (p \cdot \tilde{q})
             + \eta^2 \frac{\tilde{q}^2}{4}
           \right\}
        \right]\, ,
  \label{eq:ex_H}
\end{eqnarray}
where $\beta = \alpha_0 + \alpha_+ + \alpha_-$
and $\tilde{q}^\mu = C^{\mu\nu} q_\nu$
has the dimension of length.
 $H(0,p,q)$ is $\frac{1}{16\pi^2 m^2}$ for on-shell photon.
 If the naive commutative limit ($C^{\mu\nu}$ $\rightarrow$ 0)
were taken before integration
two contributions (the second line in eq. (\ref{eq:MDM_1}))
from Fig. \ref{fig:e_g}(b) would cancel with each other,
leaving the usual contribution to ($g-2$)
from the first graph, Fig. \ref{fig:e_g}(a).
 The calculation similar to Appendix leads
\begin{eqnarray}
 H(1,p,q)
  &=&
   \frac{1}{8\pi^2}
   \int_0^1 d\alpha_+ \int_0^{(1-\alpha_+)} d\alpha_-
   \frac{(\alpha_+ + \alpha_-) - (\alpha_+ + \alpha_-)^2}
        {(\alpha_+ + \alpha_-)^2 m^2
          + \alpha_+ \alpha_- (-q^2)}
    \nonumber \\
  && \qquad \qquad \qquad \qquad \qquad \quad
   \times
   e^{i(\alpha_+ + \alpha_-) (p\cdot \tilde{q})}\,
   x\,K_1(x)
   \, ,
\end{eqnarray}
where $x = (-\tilde{q}^2)
            \left\{
             (\alpha_+ + \alpha_-)^2 m^2
             + \alpha_+ \alpha_- (-q^2)
            \right\}$,
and $K_1(x)$ is a modified Bessel function of the second kind
\cite{math_formula}.
 At this stage we can justify that $q^2$ and $\tilde{q}^2$
in $H(1,p,q)$ is brought to zero
without confronting with any singularities
since $K_1(x) \sim \frac{1}{x}$ for $x\sim 0$.
 Thus for $q^2=0$ and $\tilde{q}^2=0$,
$H(1,p,q)$ becomes equal to $H(0,p,q)$.
 Therefore the magnetic dipole moment
does not change for the photon with no transverse
momentum along $(2,3)$-plane ($\tilde{q}$ = 0).
 The photon on mass shell can, of course, have the momentum
transverse along such a plane.
 Eq. (\ref{eq:MDM_1}) then exhibits the difference where
\begin{equation}
 \left. H(1,p,q)\right|_{q^2 = 0}
  = \frac{1}{8\pi^2 m}
    \int_0^1 ds (1-s) e^{is (p\cdot \tilde{q})}
    x(s) K_1(x(s))\, ,
\end{equation}
with $x(s) = (-\tilde{q}^2)m^2 s^2$.
 We note the two points.
 The first is that the third term
in eq. (\ref{eq:MDM_1}) carries the phase
which should be accompanied by the fields
with a charge opposite to the electron.
 The origin of the sign change can be easily understood
in terms of the double-line representation
of photon propagation \cite{Filk,Ishibashi_NCYM,Seiberg}.
 The second feature is that
nonzero $\tilde{q}^2$ enters
in the integrand with a combination
$x = (-\tilde{q}^2)
     \left\{
      (\alpha_+ + \alpha_-)^2 m^2 + \alpha_+ \alpha_- (-q^2)
     \right\}$,
or $(-\tilde{q}^2) (-q^2)$.
 Such a feature is also observed in the charge form factor.
 This indicates that the threshold behavior
around electron-positron pair production $q^2 \sim 4m^2$
is largely affected by nonzero $\tilde{q}^2$.
%%%%%%%%%%
%%%%%%%%%%%%%%%%%%%%%%%%%%%%%%%%%%%%%%%%%%%%%%%%%%%%%%%%%%%%%%%%
\subsection{Vacuum polarization of photon}
\label{sec:vac}
%%%%%%%%%%%%%%%%%%%%%%%%%%%%%%%%%%%%%%%%%%%%%%%%%%%%%%%%%%%%%%%%
\quad
 The calculation for the finite part
of the vertex functions will be necessary
to compute the cross section of, for instance,
$e^+ e^-$ annihilation process
in NC-QED.
 One important vertex function is
the photon vacuum polarization.
 Indeed the following one-loop calculation of it
gives us the interesting information
on the quantum mechanical dynamics
of noncommutative field theory.
 Here we first concentrate on the contributions from
Fig. \ref{fig:vac_photon}(a) $\sim$ (c),
which also exists in NCYM theory.
\\
\quad
 We first remind the appearance of
the singular term $\sim 1/ \tilde{q}^2$
found in eq. (\ref{eq:VP_4}) for one Feynman diagram
in Fig. \ref{fig:vac_photon}(c).
 The question is
whether it remains
even after summing up Fig. \ref{fig:vac_photon}(a) $\sim$ (c).
 The formula (\ref{eq:int_formula}) derived for the integrals
in Appendix enables us to extract the singular terms
$1/\tilde{q}^2$ and ${\rm ln}(\tilde{q}^2 q^2)$
from eq. (\ref{eq:VP_g_33})
\begin{equation}
 i \Pi^{\mu\nu}_{(a) + (b)} \sim
  i \frac{g^2}{16\pi^2}
   \left\{
     g^{\mu\nu} \frac{24}{-\tilde{q}^2}
     +
     \frac{10}{3}
     \left(
      g^{\mu\nu} q^2 - q^\mu q^\nu
     \right) {\rm ln} (q^2 \tilde{q}^2)
     + 32 \frac{\tilde{q}^\mu \tilde{q}^\nu}{\tilde{q}^4}
     - \frac{4}{3} \frac{q^2}{\tilde{q}^2}
       \tilde{q}^\mu \tilde{q}^\nu
   \right\} \, .
    \label{eq:VP_g_33_s}
\end{equation}
 In the sum of (\ref{eq:VP_4}) and (\ref{eq:VP_g_33_s})
$g^{\mu\nu} /\tilde{q}^2$ term cancels out
\begin{equation}
 i \Pi^{\mu\nu}(q) \sim
 i \frac{g^2}{16\pi^2}
   \left\{
     \frac{10}{3}
     \left(
      g^{\mu\nu} q^2 - q^\mu q^\nu
     \right) {\rm ln} (q^2 \tilde{q}^2)
     + 32 \frac{\tilde{q}^\mu \tilde{q}^\nu}{\tilde{q}^4}
     - \frac{4}{3} \frac{q^2}{\tilde{q}^2}
       \tilde{q}^\mu \tilde{q}^\nu
   \right\} \, ,
    \label{eq:singular}
\end{equation}
which is consistent with Slavnov-Taylor identity
derived from BRST symmetry.
 It should be recalled that
the nonplanar contribution would diverge
if the integral in eq. (\ref{eq:VP_g_33})
were evaluated with $\tilde{q}^2$ set equal to zero.
 The logarithmic infrared singularity ${\rm ln}(\tilde{q}^2)$
in (\ref{eq:singular}) reflects
the fact that UV divergence is at most logarithmic.
 In fact the coefficient $10/3$ of ${\rm ln}(\tilde{q}^2)$
is that of the wave function renormalization factor
(\ref{eq:Z_A_NCYM}) of photon in NCYM theory.
 Thus the term singular in $\tilde{q}^2$
correlates with the ultraviolet divergence.
 Such a phenomenon is also observed in Ref. \cite{Bigatti,Seiberg}.
 The other terms proportional to $\tilde{q}^\mu \tilde{q}^\nu$
is interesting.
 The effect of such terms is obscure until the concrete cross section
is estimated.
\\
\quad
 NC-QED has one extra contribution from Fig. \ref{fig:vac_photon} (d).
 It is entirely planar as was discussed
in Sec. \ref{sec:vac}.
 Thus the logarithmic infrared singularity
in the one-loop correction to photon vacuum polarization
of NC-QED theory
is completely the same as in NCYM theory at one-loop level.
 On the other hand the wave function renormalization
of photon receives a net effect from such a planar contribution
and results in eq. (\ref{eq:Z_A}).
 Thus NC-QED theory does not have the infrared-UV correspondence
as found in NCYM theory.
%%%%%%%%%%%%%%%
%%%%%%%%%%%%%%%%%%%%%%%%%%%%%%%%%%%%%%%%%%%%%%%%%%%%%%%%%%%%%%%%%
\section{Weyl Fermions and Chiral Gauge Theory}
\label{sec:chiral}
%%%%%%%%%%%%%%%%%%%%%%%%%%%%%%%%%%%%%%%%%%%%%%%%%%%%%%%%%%%%%%%%%
%%%%%%%%%%%%%%%
 Until now all the fermions are
assumed to be Dirac fermions.
 It is naturally tempted
to pursue the extension to chiral gauge theory.
 Since the classical analysis given in Sec. \ref{sec:basic}
is irrelevant to the chiral property of fermion,
Weyl fermions can have the charge $+1$ or $-1$.
 The right-handed fermion with $+1$ charge
is easily seen to be replaced by its CP conjugate
(the left-handed) fermion also in the present context.
 Thus the chiral gauge theory
simply implies that
the number of the left-handed fermions with $+1$ charge
is not equal to that with $-1$
\footnote{
 We require that
the triangular loop contribution cancels with each other
for {\it all} momentum configuration.
 But it might be too strong requirement
for noncommutative SU(N) gauge theory
due to non-factorizability of color and phase factors,
as suggested by Y. Kitazawa.
}.
 The question is whether such a theory
circumvents a triangular loop anomaly
to define a consistent quantum theory or not.
\\
\quad
 Let us imagine one triangular diagram
in which the fermion number flows from 1 to 2
with three incoming external momentum $q_i$ ($i=1,2,3$).
 Then the momentum conservation gives
$q_i$ in terms of the internal momenta
\begin{equation}
 q_1 = k_2 - k_3, \quad q_2 = k_1 - k_2, \quad q_3 = k_2 - k_3
  \, .
   \label{eq:eq}
\end{equation}
 The phase factor associated with the diagram in our mind is
\begin{equation}
 \exp
   \left[
    \frac{i}{2}
    \left(
     k_3\cdot C\cdot k_2 + k_2\cdot C\cdot k_1
      + k_1\cdot C\cdot k_3
    \right)
   \right]\, .
\end{equation}
 With help of eq. (\ref{eq:eq})
this phase factor reduces
to $e^{-\frac{i}{2} q_1\cdot C\cdot q_2}$.
 Thus a triangular fermion loop diagram
gives only planar contributions.
 Once we remind the correspondence between the current theory
to ordinary nonabelian gauge system
in which the external momentum
plays the role of color in the gauge theory side,
the remained integral
is evaluated in the same manner
as in ordinary nonabelian gauge theory
which involves
the fundamental and/or anti-fundamental Weyl fermions.
 From this observation,
the number of the left-handed fermions with
$-1$ charge match with the number of $+1$ in the system.
 Such a theory is vector-like, i.e.,
nothing but NC-QED
considered until the previous sections.
%%
%%%%%%%%%%%%%%%%%%%%%%%%%%%%%%%%%%%%%%%%%%%%%%%%%%%%%%%%%%%%%%%%%
\section{Summary}
\label{sec:conc}
%%%%%%%%%%%%%%%%%%%%%%%%%%%%%%%%%%%%%%%%%%%%%%%%%%%%%%%%%%%%%%%%%
%%
\quad
 In this paper NC-QED,
which is the simplest extension of
U(1) NCYM theory,
is studied to observe the UV structure
and the perturbative aspects of the low momentum region.
 It is observed that
one loop diagram are all made finite
by the redefinition of the operators
and parameters originally involved in
a classical action.
 From the evaluation of those one loop diagrams,
the UV divergence is suggested to appear only
in the planar diagram,
which is the same feature shared already in
NCYM theory.
 Although NCYM theory
accommodates SU(2)-like structure in UV divergence,
U(1) facet appears in $\beta$ function when
the electron is introduced.
\\
\quad
 The possibility of the extension to chiral gauge theory
is examined, but the simple extension does not admit
any chiral gauge theory.
\\
\quad
 To observe the infrared behavior,
the anomalous magnetic dipole moment
is evaluated in this theory.
 The value is the same as in conventional QED
for the photon propagating in the direction
orthogonal to the plane on which noncommutativity enters.
 In the transverse direction,
the magnetic coupling depends on
the magnitude of the components
of the momentum along this direction.
 This exhibits the violation of full Lorentz invariance SO(1,3)
through the breaking parameter $C^{23}$.
 It is desirable to know the next-order correction
which will pick up the one-loop modification
to the photon propagator.
 Also the explicit evaluation of
the cross section of $e^+ e^-$ annihilation process,
Compton scattering process, etc.,
should be carried out to capture the relationship
with this explicit violation more directly.
\\
\quad
 The finite part of the photon vacuum polarization
indicates that the singular terms are correlated
with UV divergent structure,
which shows up the essential aspects
of the quantum mechanical dynamics of NCYM theory
not shared by NC-QED theory.
 The similar analysis for NCYM on the finite volume
torus is interesting to further clearly the point,
as was done for the analysis of UV divergence
\cite{J_NCYM,Krajewski}.
  It is also important to know the special role of
of supersymmetry which can control UV structure
in the noncommutative system
which is naturally obtained in the context
of superstring theory \cite{MLi}.
 The above connection of UV and infrared limit
invokes us about the string theory together with
the connection of gauge theory to the interacting string theory
\cite{tHooft,Polyakov,Matrix}.
 The problem is that
the perturbative string theory is absent
of UV divergence as insured by its moduli invariance
while the noncommutative theory has UV divergence generally.
 Then supersymmetry will play the key role.
 Such a subject should be further investigated similarily
to Ref. \cite{Seiberg} attempting to construct
the string theory which does not refer
to any world-sheet points of view.
\\
\\
\quad
\\
{\bf \Large Acknowledgements}
\quad
\\
\\
\quad
 The author thanks greatly
S. Iso for discussion and suggestion
at frequent times and reading manuscript.
 He also thanks L. Susskind for pointing out the serious error
in the result at the stage of Ref. \cite{Hayakawa},
and N. Ishibashi, Y. Kitazawa, K. Okuyama
and F. Sugino
for learning about noncommutative theory,
and the colleagues at KEK for sharing common interests
in this and other various topics
at a weekly informal meeting.
%%
%%%%%%%%%%%%%%%%%%%%%%%%%%%%%%%%%%%%%%%%%%%%%%%%%%%%%%%%%%%%%%%%%
%%%%%%%%%%%%%%%%%%%%%%%%%%%%%%%%%%%%%%%%%%%%%%%%%%%%%%%%%%%%%%%%%
%\appendix
%\section{Appendix}
%\label{sec:app_A}
\\ \quad
\\ \quad
\\ %\vspace{0.5cm}
{\bf \Large Appendix}
%%%%%%%%%%%%%%%%%%%%%%%%%%%%%%%%%%%%%%%%%%%%%%%%%%%%%%%%%%%%%%%%%
%%%
\\
\quad
\\
\quad
 Here to illustrate the convergence
of the various nonplanar portions of diagrams,
the ghost loop contribution in Fig. \ref{fig:vac_photon}(a)
to the vacuum polarization
of the gauge field is explicitly evaluated.
 The application of Feynman rule in Fig. \ref{fig:Feynman_rule}
gives to the associated contribution
under the dimensional regularization
\begin{eqnarray}
 i\Pi^{\mu\nu}_{\rm gh}(q) &=&
  -i(2g)^2\int \frac{d^d k}{i(2\pi)^d}
  \sin^2 \left( \frac{1}{2} k \cdot C \cdot q \right)
  \frac{(k-q/2)^\mu (k+q/2)^\nu}
       {\left((k-q/2)^2 + i\epsilon\right)
        \left((k+q/2)^2 + i\epsilon\right)}\, .
    \label{eq:ghost_i}
\end{eqnarray}
 Schwinger parameterization of the propagator \cite{Itzykson}
\begin{equation}
 \frac{1}{k^2 - m^2 + i\epsilon}
  = -\int_0^\infty id \alpha e^{i\alpha(k^2 - m^2 + i\epsilon)}
   \, ,
\end{equation}
will enable us
to carry out
the integration over loop momentum $k$ including the phase factor.
 Eq. (\ref{eq:ghost_i}) gives
\begin{eqnarray}
 i\Pi^{\mu\nu}_{\rm gh}(q) &=&
  -i \frac{(2g)^2}{2}
  \int_0^\infty id \alpha_+ \int_0^\infty id \alpha_-
   \left[ I^{\mu\nu}(0,q) -
          \frac{1}{2} \left(
                        I^{\mu\nu}(1,q) + I^{\mu\nu}(-1,q)
                      \right)
   \right]
   \, ,
   \nonumber \\
\end{eqnarray}
where
\begin{eqnarray}
 I^{\mu\nu}(\eta,q) &=&
  p_-^\mu p_+^\nu \int \frac{d^d k}{i(2\pi)^d}
  \exp\left[
       i\left\{
          \alpha_+ \left(k+\frac{q}{2}\right)^2 +
          \alpha_- \left(k-\frac{q}{2}\right)^2
        \right.
       \right.
   \nonumber \\
 && \quad \quad \quad \quad \quad \quad \quad \quad
   \left.
       \left.
        \left.
          + \eta k\cdot \tilde{q}
          + z_+ \cdot \left( k+\frac{q}{2}\right)
          + z_- \cdot \left( k-\frac{q}{2}\right)
        \right\}
       \right]
   \right|_{z_\pm \rightarrow 0}
    \, ,
\end{eqnarray}
with $p^\mu_{\pm} =\frac{1}{i} \frac{\del}{\del z_{\pm\,\mu}}$
is a derivative operator and
$\tilde{q}^\mu = C^{\mu\nu} q_\nu$.
 Then the integral can be brought into the form of Gaussian-type.
 Use of
\begin{equation}
 \int \frac{d^d k}{i(2\pi)^d} e^{i\beta k^2}
  = \frac{1}{(4\pi\beta i)^{d/2}}\, ,
\end{equation}
and performing the derivative operation yields
\begin{eqnarray}
 I^{\mu\nu}(\eta,q) &=&
  \frac{1}{(4\pi\beta i)^{d/2}}
    \exp\left[
        -i
        \left\{
         \frac{\alpha_+ \alpha_-}{\beta} (-q^2)
         + \frac{1}{\beta} \frac{\eta^2 \tilde{q^2}}{4}
        \right\}
      \right]
    \nonumber \\
 && \quad \times
  \left[
   \frac{i}{2\beta} g^{\mu\nu}
   - \frac{\alpha_+ \alpha_-}{\beta^2} q^\mu q^\nu
   + \frac{\eta^2}{4} \tilde{q}^\mu \tilde{q}^\nu
   + ({\rm terms\ linear\ in\ \eta})
  \right] \, ,
\end{eqnarray}
where
$\beta$ is $\beta = \alpha_+ + \alpha_-$.
 Thus the expression for the ghost loop contribution
becomes
\begin{eqnarray}
 i\Pi^{\mu\nu}_{\rm gh}(q) &=&
  -i \frac{(2g)^2}{2} \int_0^\infty i d\alpha_+
     \int_0^\infty i d\alpha_- \frac{1}{(4\pi\beta i)^{d/2}}
     \exp\left(
           -i \frac{\alpha_+ \alpha_-}{\beta} (-q^2)
         \right)
   \nonumber \\
 && \quad \quad \times
  \left\{
   \left(
    1 - e^{-\frac{i}{\beta} \frac{\tilde{q}^2}{4}}
   \right)
   \left(
    \frac{i}{2\beta} g^{\mu\nu}
    - \frac{\alpha_+ \alpha_-}{\beta^2} q^\mu q^\nu
   \right)
   -
   e^{-\frac{i}{\beta} \frac{\tilde{q}^2}{4}}
   \frac{1}{4\beta^2} \tilde{q}^\mu \tilde{q}^\nu
  \right\} \, .
   \label{eq:gh_2}
\end{eqnarray}
 The term not proportional to
$e^{-\frac{i}{\beta}\frac{\tilde{q}^2}{4}} $ in the bracket
of eq. (\ref{eq:gh_2}) gives a planar contribution
and is UV divergent.
\\
\quad
 The nonplanar contributions in (\ref{eq:gh_2}),
on the other hand, are convergent.
 Indeed the first term with nontrivial phase
in eq. (\ref{eq:gh_2}) becomes
for $d=4$ ($\tilde{q}$ is space-like four momentum.)
\begin{eqnarray}
 &&
 \int_0^\infty id\alpha_+  \int_0^\infty id\alpha_-
  \frac{1}{(4\pi\beta i)^{d/2}} \frac{1}{\beta}
  \exp
   \left[
    -i
    \left\{
     \frac{\alpha_+ \alpha_-}{\beta} (-q^2)
     + \beta \mu^2
     + \frac{1}{\beta} \frac{\tilde{q}^2}{4}
    \right\}
   \right]
    \nonumber \\
 && \quad =
 i \frac{1}{16\pi^2}
  \int_0^1 d\alpha_+
   \left\{ \alpha_+(1-\alpha_+)(-q^2) + \mu^2 \right\}
   \nonumber \\
 && \qquad \qquad \quad \quad \times
  \int_0^\infty \frac{d\rho}{\rho^2}
  \exp\left[
        -\rho
        - \frac{1}{\rho}
           \left(
            \frac{-\tilde{q}^2}{4}
           \right)
           \left\{
            \alpha_+ (1-\alpha_+)(-q^2) + \mu^2
           \right\}
      \right]\, ,
  \label{eq:int_1}
\end{eqnarray}
where the small mass $\mu$ for the FP-ghost
is introduced to regularize the infrared divergence,
which will be found unnecessary hereafter.
 The UV behavior is characterized by
the contribution from the integral around $\rho \sim 0$.
 Since the integral around the lower end in terms of
$\lambda = 1/\rho$ becomes
\begin{equation}
 \int_0^a \frac{d\rho}{\rho^n}
       \exp \left(-\rho - \frac{1}{\rho} a^2 \right)
 =
 \int_{1/a}^\infty d\lambda \lambda^{n-2}
  \exp\left(-a^2 \lambda - \frac{1}{\lambda} \right)
 \, ,
\end{equation}
nonzero $\tilde{q}^2$ avoids
the integral to diverge around $\rho \sim 0$
($\lambda \rightarrow \infty$)
for any power-like singularity.
 Actually the modified Bessel function of the second order
writes the integral for $a > 0$
\begin{equation}
 \int_0^\infty \frac{d\rho}{\rho^{n+1}}
  \exp\left( -\rho - \frac{1}{\rho} a^2 \right)
 = \left( -\frac{1}{2a} \frac{d}{da} \right)^n [2K_0(2a)]\, .
   \label{eq:int_Bessel}
\end{equation}
 The asymptotic expansion of $K_0(x)$ around $x\sim0$
is known \cite{math_formula} as
\begin{eqnarray}
 K_0(2a) &=&
  \sum_{k=0}^{\infty} \frac{a^{2k}}{(k !)^2}
   \left(
     - {\rm ln}(a) + \psi(k+1)
   \right) \, ,
    \label{eq:asy_Bessel}
\end{eqnarray}
where $\psi(z) = d{\rm ln}\Gamma(z) /dz$ becomes
for an integer $z=n$
\begin{equation}
 \psi(1) = - \gamma_E, \quad
 \psi(n) = -\gamma_E + \sum_{k=1}^{n-1} \frac{1}{k}
  \quad (n\ge 2)\, ,
\end{equation}
with Euler constant $\gamma_E$.
 Eqs. (\ref{eq:int_Bessel}) and (\ref{eq:asy_Bessel}) give
the asymptotic expansion for the integrals entering in our
one-loop calculus
\begin{eqnarray}
 \int_0^\infty \frac{d\rho}{\rho}
 \exp\left( -\rho - \frac{1}{\rho} a^2 \right)
 &=&
  -{\rm ln}(a^2)
     \left( 1 + a^2 + {\cal O}(a^4) \right)
  - 2 \gamma_E + (-2\gamma_E + 2) a^2
   \nonumber \\
 && \quad
  + {\cal O}(a^4)\, ,
  \nonumber \\
 \int_0^\infty \frac{d\rho}{\rho^2}
 \exp\left( -\rho - \frac{1}{\rho} a^2 \right)
 &=&
  {\rm ln}(a^2)
   \left(
    1 + \frac{1}{2} a^2 + {\cal O}(a^4)
   \right)
   \nonumber \\
 && \quad
  + \frac{1}{a^2}
  +
  \left(
   2 \gamma_E - 1
  \right)
  + \left( \gamma_E - \frac{5}{4} \right) a^2
  + {\cal O}(a^4)\, ,
   \nonumber \\
 \int_0^\infty \frac{d\rho}{\rho^3}
  \exp\left( -\rho - \frac{1}{\rho} a^2 \right)
 &=&
  -{\rm ln}(a^2)
   \left(
    \frac{1}{2} + \frac{1}{6} a^2 + {\cal O}(a^4)
   \right)
   + \frac{1}{a^4} - \frac{1}{a^2}
     \nonumber \\
 && \quad 
   + \left(
      -\gamma_E + \frac{3}{4}
     \right)
   + \left(
      -\frac{1}{3} \gamma_E + \frac{39}{36}
     \right) a^2 + {\cal O}(a^4)\, .
      \label{eq:int_formula}
\end{eqnarray}
where $n=2$ to use for eq. (\ref{eq:int_1}).
%
%%%%%%%%%%%%%%%%%%%%%%%%%%%%%%%%%%%%%%%%%%%%%%%%%%%%%%%%%%%%%%%%
%

%
%
\end{document}